\documentclass[aps, twocolumn, prd, preprintnumbers, amsmath, amssymb, amsfonts, superscriptaddress, nofootinbib]{revtex4-1}
\pdfoutput=1

\usepackage{bm, physics, mathtools}
\usepackage{enumitem}

\usepackage[pagebackref=false,hidelinks]{hyperref}

\usepackage{setspace,latexsym}
\usepackage{color}
\usepackage{epsfig}
\usepackage{graphicx}
\usepackage{slashed}
\usepackage[export]{adjustbox}
\usepackage{cancel}

\newcommand{\beq}{\begin{equation}}
\newcommand{\eeq}{\end{equation}}
\newcommand{\bea}{\begin{eqnarray}}
\newcommand{\eea}{\end{eqnarray}}
\newcommand{\nn}{\nonumber}

\newcommand{\eV}{\mathrm{eV}}

\newcommand{\GeV}{\mathrm{GeV}}
\newcommand{\MeV}{\mathrm{MeV}}
\newcommand{\TeV}{\mathrm{TeV}}

\newcommand{\thdot}{\dot{\theta}}
\newcommand{\mD}{m_\Delta}

\newcommand{\nEq}[1]{n^{\rm eq}_{#1}}
\def\bal#1\eal{\begin{align}#1\end{align}}

\begin{document}

\title{Majoron-driven spontaneous leptogenesis in type-II seesaw model
}

\author{Eung Jin Chun}
\email{ejchun@kias.re.kr}
\affiliation{Korea Institute for Advanced Study, Seoul 02455, South Korea}

\author{Tae Hyun Jung}
\email{thjung0720@gmail.com}
\affiliation{Particle Theory  and Cosmology Group, Center for Theoretical Physics of the Universe,
Institute for Basic Science (IBS),
 Daejeon, 34126, Korea}

\author{Jin Sun}
\email{jinsun930503@gmail.com}
\affiliation{Particle Theory  and Cosmology Group, Center for Theoretical Physics of the Universe,
Institute for Basic Science (IBS),
 Daejeon, 34126, Korea}
 \affiliation{State Key Laboratory of Dark Matter Physics, Tsung-Dao Lee Institute \& School of Physics and Astronomy,
Shanghai Jiao Tong University, Shanghai 200240, China}

\preprint{CTPU-PTC-26-22}

\begin{abstract}
We investigate a minimal realization of spontaneous leptogenesis in the type-II seesaw model, where the lepton and Higgs asymmetries are generated as decay and inverse-decay processes drive the plasma toward the displaced equilibrium dictated by a rotating Majoron background. 
We derive the complete set of Boltzmann equations governing the evolution of asymmetric particle densities and analytically and numerically identify the freeze-out and freeze-in regimes. 
The resulting asymmetries depend primarily on the triplet branching fractions and are nearly independent of the triplet mass. 
The asymmetry scales as the square root of the smaller branching fraction when either decay mode into leptons or Higgs fields is suppressed, reflecting the symmetry structure; the lepton number is conserved once either coupling is turned off. 
We also show that the Majoron background does not induce CP-asymmetric decay of the scalar triplet in the massless final state limit, and thus its impact is negligible. 
Our results establish spontaneous leptogenesis as a simple and robust alternative to thermal type-II leptogenesis that requires neither additional scalar triplets nor explicit CP-violating interactions. 
\end{abstract}

\maketitle

\section{Introduction}
\label{sec:intro}

The origin of the baryon asymmetry of the Universe, $\eta_B = n_B/n_\gamma \simeq 6.1\times 10^{-10}$ as inferred from the cosmic microwave background (CMB) and big-bang nucleosynthesis (BBN)\,\cite{Planck:2018vyg}, remains one of the central open questions of particle cosmology. 
Any dynamical explanation must satisfy the three Sakharov conditions---baryon number violation, $C$ and $CP$ violation, and a departure from thermal equilibrium\,\cite{Sakharov:1967dj}---and the Standard Model (SM) alone fails to provide sufficient $CP$ violation and a strong enough departure from equilibrium.

Leptogenesis \cite{Fukugita:1986hr} is arguably the most economical solution: a lepton asymmetry produced by the $CP$-violating, out-of-equilibrium decays of heavy states is partially converted into a baryon asymmetry by electroweak sphaleron processes\,\cite{Kuzmin:1985mm}, tying baryogenesis directly to the origin of neutrino masses. 
In the type-II seesaw realization\,\cite{Magg:1980ut, Schechter:1980gr, Cheng:1980qt, Lazarides:1980nt, Mohapatra:1980yp}, neutrino masses arise from the induced vacuum expectation value (vev) of an $SU(2)_L$-triplet scalar $\Delta$. 
Thermal leptogenesis in this framework, however, faces well-known obstructions: with a single triplet, the tree-level and one-loop decay amplitudes cannot interfere to produce a $CP$ asymmetry, so that at least two triplets or additional heavy states are required\,\cite{Ma:1998dx, Hambye:2003ka, Hambye:2005tk, Chun:2006sp}; moreover, the gauge interactions of the triplet keep its abundance close to equilibrium and tend to suppress the efficiency\,\cite{Hambye:2005tk, Chun:2006sp}.

Spontaneous baryogenesis/leptogenesis\,\cite{Cohen:1987vi, Cohen:1988kt} offers a qualitatively different route. 
If a scalar field couples derivatively to a particle-number current, a nonvanishing velocity $\thdot=d\theta/dt$ of its phase $\theta$ acts as an effective chemical potential that splits the energies of particles and antiparticles, biasing number-violating interactions. 
The asymmetry is thus generated without $CP$-violating out-of-equilibrium decays, the required departure from equilibrium being supplied by the rolling background itself, which effectively violates $CPT$. 
This idea has been realized with axions and axion-like fields\,\cite{Kusenko:2014uta, Co:2019wyp, Co:2020xlh, Co:2020jtv, Domcke:2020kcp, Harigaya:2021txz, Co:2021qgl, Chakraborty:2021fkp, Kawamura:2021xpu, Mukaida:2021sgv, Co:2021qgl, Co:2022aav, Barnes:2022ren, Co:2022kul, Badziak:2023fsc, Berbig:2023uzs, Barnes:2024jap, Datta:2024xhg, Kuckenberg:2026oax}, and, most relevant to
us, with the majoron\,\cite{Ibe:2015nfa, Chun:2023eqc, Chao:2023ojl, Chun:2024gvp, Wada:2024cbe, Berbig:2025hlc, Datta:2026aks}---the pseudo-Nambu-Goldstone boson of spontaneously broken lepton number---whose kinetic motion can drive leptogenesis through lepton-number-violating decays and inverse decays. 
A quantitative Boltzmann analysis of this mechanism in the type-I seesaw has recently been carried out in Ref.\,\cite{Chun:2025abp}, where the right-handed neutrino Yukawa interactions equilibrate the asymmetry toward the value dictated by $\thdot$.
In this work, we explore the majoron-driven spontaneous leptogenesis scenario in the type-II see-saw model, which was first studied in Ref.\,\cite{Berbig:2025hlc}. 
Here, we construct a set of complete and consistent Boltzmann equations from first principles, and provide analytical and numerical solutions in the entire range of the parameter space.

A nonvanishing $\thdot$ itself calls for a dynamical origin. 
The most widely studied possibility is the kinetic misalignment mechanism\,\cite{Co:2019jts, Chang:2019tvx}: if an explicit breaking of the global symmetry is operative when the radial mode takes a large field value in the early Universe---as in the Affleck-Dine mechanism\,\cite{Affleck:1984fy}---the angular mode receives a kick, and the field starts to rotate in field space. 
Since the charge density of the rotating field $n_\theta = \thdot f_\theta^2$ with $f_\theta$ being the field value of the radial mode, is conserved in a comoving volume, the velocity subsequently redshifts as $\thdot \propto a^{-3} \propto T^3$ for the scale factor $a$.
Alternative sources of $\thdot$ include a conventional (potential) misalignment\,\cite{Cohen:1987vi, Cohen:1988kt, Ibe:2015nfa}, effective $CPT$-violating backgrounds such as
torsion\,\cite{deCesare:2014dga}, electroweak first-order phase transition\,\cite{Jeong:2018ucz, Jeong:2018jqe, Im:2021xoy, Harigaya:2023bmp, Jeong:2024hhi, Bhandari:2025phe, Wang:2026jjn}, and sliding generated with the help of a symmetry non-restoration mechanism\,\cite{Chun:2024gvp}.
In this work, we assume that the majoron velocity is generated by the kinetic misalignment mechanism, parametrizing it by $\epsilon_1 \equiv (\thdot/T)|_{T=\mD}$, while leaving the details undetermined.

In our scenario, both decay channels of $\Delta$ to leptons ($\Delta \to \ell \ell$) and Higgs fields ($\Delta \to HH$) are essential. 
If either their branching ratios, denoted by ${\rm Br}_\ell$ and ${\rm Br}_H$, vanish, the $B-L$ charge of $\Phi$ can be reassigned so that the remaining interactions are exactly $U(1)_{B-L}$ symmetric; $\theta$ is then unphysical, and no asymmetry can be generated. 
The final asymmetry must therefore vanish in both limits, and we find numerically and analytically that it is suppressed as $\sqrt{{\rm Br}_\ell {\rm Br}_H}$ whenever either branching ratio is hierarchically small, while the yield is essentially independent of the triplet mass. 
We derive the full set of Boltzmann equations for the asymmetric yields, including the $\thdot$-induced source terms, solve them numerically, and provide analytic estimates of the freeze-in and freeze-out behavior in each regime.

One might expect a difference between particle and antiparticle reaction rates in the background of $\thdot$.
In general, this is true for reactions that violate the broken $U(1)$ symmetry corresponding to the $\theta$ shift symmetry, e.g., $\Delta \to H H$ in our scenario.
However, when final-state particle masses are small, the difference is suppressed. 
We show that in the massless-Higgs limit the shifts of energies cancel inside the phase-space integrals, and thus $\Gamma_{\Delta \to HH}-\Gamma_{\bar \Delta \to \bar H \bar H}$ is at $O(m_H^2 \thdot/m_\Delta^3)$, which is negligible when $m_\Delta \gg m_H$.  
This should be contrasted with the type-I scenario, in which a heavy Majorana neutrino undergoes CP-asymmetric decays at tree level \cite{Chun:2025abp}.

What $\thdot$ does modify is the equilibrium state toward which the system relaxes. 
The distribution functions shifted by nonzero $\thdot$ carry nonzero chemical potentials fixed by hypercharge neutrality together with the chemical-equilibrium conditions, and the asymmetry is generated as the plasma is driven toward this shifted equilibrium. 
The obstacles in the thermal leptogenesis case do not occur in our scenario; the large gauge annihilation rate is simply irrelevant in the generation of asymmetric abundance, and the strong washout factor enforced by the neutrino mass scale is exactly what guarantees an efficient production of the asymmetric components toward $\thdot$-shifted equilibrium values. 
Unlike thermal leptogenesis in the type-II seesaw, no second triplet is required.

The paper is organized as follows. 
In Sec.\,\ref{sec:model} we introduce the type-II seesaw model with majoron and its connection to neutrino masses. 
Section\,\ref{sec:dispersion} derives the dispersion relations in the $\thdot$ background, and Sec.\,\ref{sec:decay} the modified decay rates of $\Delta$. 
Section\,\ref{sec:boltzmann} presents the Boltzmann equations, and numerical results are given in Sec.\,\ref{sec:results}.
Relic majoron abundance is estimated in Sec.\,\ref{sec:cogenesis}, and we conclude in Sec.\,\ref{sec:conclusion}.

\section{Majoron in type-II seesaw model}
\label{sec:model}

In the type-II seesaw mechanism, an $SU(2)_L$-triplet complex scalar $\Delta$ with hypercharge $Y = -1$ is introduced, whose Yukawa coupling to the lepton doublets generates neutrino masses once $\Delta$ acquires an induced vev. 
We extend the model by a complex singlet $\Phi$ and impose a global $U(1)_{B-L}$ symmetry under which the lepton doublet $\ell$, Higgs $H$, $\Delta$, and $\Phi$ carry the charges $-1$, $0$, $-2$ and $2$, respectively.
The interactions relevant for our discussion are
\bal
  -{\cal L} \supset
  \frac{1}{2} y_{\alpha\beta} \bar{\ell}_\alpha^c \ell_\beta \Delta^\dagger
  + \frac{1}{2} \lambda_H H H \Delta^\dagger \Phi^* + \text{h.c.},
  \label{eq:lagrangian}
\eal
together with the scalar sector potential
\bal
V \supset m_\Delta^2 |\Delta|^2 
+ \lambda_\Phi \Big(|\Phi|^2-\frac{v_\sigma^2}{2} \Big)^2 + \lambda_{\Phi \Delta} |\Phi|^2 |\Delta|^2 \cdots,
\label{eq:potential}
\eal
where the ellipsis includes the SM Higgs potential, and the remaining quartic terms that play no role in what follows.

An important consequence of the charge assignment is that the bare trilinear interaction $HH\Delta^\dagger$ of the conventional type-II seesaw carries $B-L=2$ and is therefore forbidden. 
It arises only after $U(1)_{B-L}$ is spontaneously broken by the vev of $\Phi$. 
Writing
\bal
  \Phi = \frac{v_\sigma + \sigma}{\sqrt{2}}\, e^{i\theta},
  \label{eq:ssb}
\eal
the effective trilinear interaction $\mu_H \Delta^\dagger HH$ is generated with 
\bal
  \mu_H = \frac{\lambda_H v_\sigma}{2\sqrt{2}},
\eal
and the angular mode $\theta$ is identified with the majoron $J=\theta \,v_\sigma$. 
In this construction, the smallness of $\mu_H$, and hence of the induced triplet vev, is controlled by the symmetry-breaking scale $v_\sigma$ and the quartic coupling $\lambda_H$.

Once $\Phi$ acquires its vev, the induced trilinear interaction is the only interaction violating lepton number in the effective theory below $v_\sigma$, and one might therefore take $\Delta \to HH$ alone to be the process generating the lepton asymmetry.
This is not the case: both $y_{\alpha \beta}$ and $\lambda_H$ must be nonzero for the mechanism to operate. 
If $y_{\alpha \beta} = 0$, lepton number survives as an independent and exactly conserved $U(1)$ under which neither $\Delta$ nor $\Phi$ is charged, so that the spontaneous breaking by $v_\Phi$ has nothing to do with it.
Needless to say, if $\lambda_H = 0$, $\Phi$ decouples from the visible sector altogether. 
In either limit $\theta$ can be removed by a field redefinition and cannot bias lepton number, however fast it rolls. 
Both $\Delta \to \ell \ell$ and $\Delta \to HH$ are thus involved, and the final asymmetry vanishes when either coupling --- equivalently, either branching ratio --- goes to zero.

After electroweak (EW) symmetry breaking, the trilinear coupling induces a small vev for the neutral component of the triplet,
\bal
  v_\Delta & \equiv 
  \langle \Delta^0 \rangle = \mu_H \frac{v_{\rm EW}^2}{\mD^2},
  \qquad
  m_{\nu, \alpha \beta} = y_{\alpha \beta}\,v_\Delta
  \label{eq:numass}
\eal
where $v_{\rm EW} = \langle H^0 \rangle = 174\,\GeV$ is the SM Higgs vev, and $\mD= \sqrt{M_\Delta^2 +\lambda_{\Phi \Delta} v_\sigma^2/2}$ is the triplet mass.

We call the Goldstone mode of the $U(1)_{B-L}$ breaking as Majoron\,\cite{Chikashige:1980ui, Gelmini:1980re, Choi:1991aa} $J\simeq v_\sigma \theta$.
More precisely, since both $v_\Phi$ and $v_\Delta$ break it, the Majoron is a mixture of $\Im \Phi$ and $\Im \Delta$, where the $\Im \Delta$ contribution is suppressed by $v_\Delta/v_\sigma$ and we ignore it.

The Majoron is exactly massless at the level of Eqs.\,\eqref{eq:lagrangian} and \eqref{eq:potential}. 
We assume that an explicit breaking responsible for the initial kick of the angular mode is operative only at large radial field values, as in the Affleck–Dine mechanism, and is entirely negligible at the temperatures relevant for asymmetry generation. 
The Majoron therefore behaves as a free, massless field throughout our analysis.

As will be shown later, our leptogenesis scenario is essentially independent of $m_\Delta$ if it is greater than the electroweak phase transition temperature, and thus it is natural to see how low $m_\Delta$ may be pushed phenomenologically.
Current LHC searches impose the following bounds on the triplet mass,
depending on the dominant decay mode of the doubly charged component:
\begin{align}
  y \gg \frac{\mu_H}{m_\Delta}:\;\;
  &\mathrm{Br}(\Delta^{++}\to\ell^+\ell'^+)\sim \frac{1}{6},
  \nonumber\\
  &\mD > 1080~\mathrm{GeV}~\text{\cite{ATLAS:2022pbd}},
  \nonumber\\
  y \ll \frac{\mu_H}{m_\Delta}:\;\;
  &\Delta^{++}\to W^+W^+,
  \nonumber\\
  &\mD > 350~\mathrm{GeV}~\text{\cite{ATLAS:2021jol,Ashanujjaman:2021txz}},
  \label{eq:lhc}
\end{align}
while Ref.\,\cite{Ghosh:2026vqx} reanalyzed the leptonic channel and obtained a weaker bound $m_\Delta \gtrsim 950\,\GeV $.
In the large Yukawa limit $y\gg \mu_H/m_\Delta$,
the type-II seesaw framework offers a remarkable collider probe of neutrino mass patterns, observable through same-sign dilepton events from doubly charged boson decays \cite{Chun:2003ej}.

\section{
Triplet decays in the $\thdot$ background}

After the symmetry breaking driven by \eqref{eq:ssb}, $\theta$ dependence appears in the quartic interaction in the Lagrangian \eqref{eq:lagrangian}.
It can be removed by the field redefinition
\bal
\Delta \to \Delta\, e^{-i\theta},
\qquad 
\ell \to \ell\, e^{-i\theta/2},
\eal
without generating $\theta$ dependence in the Yukawa interactions.
In addition, in order to avoid $\theta$ dependence in the SM Yukawa interactions as well as the anomaly interactions, we also need to rotate other fermions as $\psi \to \psi \,\exp[i(B-L)_\psi \theta/2]$, where $(B-L)_\psi$ is the $B-L$ charge of $\psi$.
For simplicity, we do not include the quark sector in the following discussion.

In this basis, the effect of the rolling majoron, i.e. $\thdot \neq 0$, is contained in the modified dispersion relations of $\Delta$ and $\ell$, which we derive in Sec.\,\ref{sec:dispersion}.

The natural expectation is then that these shifts induce a difference between the decay rates of $\Delta$ and $\bar \Delta$ as in conventional CP-violating decays. 
We show in Sec.\,\ref{sec:decay} that this expectation fails: both partial widths are $\thdot$-independent in the massless final-state limit.

\subsection{Modified dispersion relations}
\label{sec:dispersion}

\subsubsection{Dispersion of $\Delta$}

After the field redefinition, the triplet kinetic term becomes
\begin{align}
  & ~~ \partial_\mu \Delta\, \partial^\mu \Delta^\dagger
  \nn \\
  &\to
  \partial_\mu \Delta\, \partial^\mu \Delta^\dagger
  +|\Delta|^2 \partial_\mu\theta \partial^\mu \theta
  - i\,\partial_\mu\theta
    \left(\Delta\, \partial^\mu \Delta^\dagger - \Delta^\dagger\partial^\mu \Delta\right), 
\label{eq:Delta_kin}
\end{align}
and the equation of motion for $\Delta$ is modified as
\bal
\big[ \partial^2 + \mD^2 - 2i \partial_\mu \theta \partial^\mu - (\partial_\mu \theta)^2 -i \partial^2 \theta \big] \Delta = 0. 
\eal
For the plane-wave mode $\Delta \sim b_\Delta e^{-i p_+\cdot x} + d^\dagger_\Delta e^{i  p_-\cdot x}$
with $p_\pm = (E_\pm, \vec p)$ and omitting the phase space integration (whose exact form under $\thdot$ background will be given later), 
the dispersion relation can be derived as
\bal
& E_\pm^2- |\vec p \,|^2 \pm 2\partial_\mu\theta p^\mu + \partial_\mu\theta \partial^\mu\theta =  m_\Delta^2, \nn
\\
&\qquad \to  E_\pm = \sqrt{|\vec p\,|^{2} + m_\Delta^2} \mp \thdot.
\label{eq:dispersion_Delta}
\eal 
We keep only the time component of $\partial_\mu\theta \to \dot \theta$, and ignore $\ddot \theta$ as $\ddot \theta \sim -3H\thdot$ is suppressed by the Hubble rate.
Here, the upper (lower) sign holds for particle (anti-particle).

The dispersion relation becomes simply $P_\pm^2 =m_\Delta^2$ if we define $P_{\pm \mu} = p_{\pm \mu} \pm \partial_\mu \theta$.
The spinor solutions of Dirac fermions obey analogous relations, which will be detailed in the following discussion.

\subsubsection{Dispersion of leptons}

Dispersion relation and spinor solutions of fermions in the $\thdot$ background have been studied in \cite{Chun:2023eqc,Chun:2025abp} for the type-I see-saw model. 
Let us recapitulate the basic features for the completeness of presentation and further study of decay processes.
In the non-linear realization (7),  the kinetic terms of the lepton sector are given by
\begin{align}
\mathcal{L}_D
&=
\bar \ell_L i \gamma^\mu \partial_\mu \ell_L
+
\bar e_R i \gamma^\mu \partial_\mu e_R
-
m \left(\bar e_L e_R + \bar e_R e_L \right) \nn \\
&\to \bar \ell_L  \gamma^\mu \left(i\partial_\mu +\frac{\partial_\mu \theta}{2}\right) \ell_L
+
\bar e_R  \gamma^\mu 
\left(i\partial_\mu +\frac{\partial_\mu \theta}{2}\right) e_R
\nn \\
& \qquad \qquad -m \left(\bar e_L e_R + \bar e_R e_L \right)
\end{align}
In terms of spinor solutions,
\begin{align}
    &\psi_L\sim u_L b e^{-i p_+ \cdot x}+ v_L d^\dagger e^{i p_- \cdot x}, 
    \nn \\
    &\psi_R\sim u_R b e^{-i p_+ \cdot x}+ v_R d^\dagger e^{i p_- \cdot x}
\end{align}
with $p_\pm = (E_\pm, \vec p)$,
we can obtain the spinor equations for $u/v$-spinors as
\begin{align}
P_{+ \mu}\bar\sigma^\mu u_L = m\, u_R, 
~&~~~~
P_{+ \mu}\sigma^\mu u_R = m\, u_L, \nn
\\
-P_{-\mu}\bar\sigma^\mu v_L = m\, v_R,
~&
-P_{-\mu}\sigma^\mu v_R = m\, v_L,
\end{align}
where $P_{\pm\mu} \equiv p_{\pm\mu} \pm \frac{1}{2}\partial_\mu \theta$. Thus, we obtain dispersion relations given as
\bal
&(P_+\cdot \sigma)(P_+\cdot \bar\sigma)=m^2 \qquad \text{for $u_L$ and $u_R$},\nn
\\
& (P_-\cdot \sigma)(P_-\cdot \bar\sigma)=m^2 \qquad \text{for $v_L$ and $v_R$,}
\eal
resulting in $P_\pm^2 = m^2$.
Therefore, taking only the time component of $\partial_\mu \theta \to \thdot$ nonzero, they lead to
\begin{align}
& 
E_\pm  = \sqrt{|\vec p\,|^{\,2}+m^2} \mp \frac{\thdot}{2},
\label{eq:dispersion_lepton}
\end{align}
where the upper (lower) sign holds for leptons (anti-leptons).

Using $P_\pm^\mu$, we write the spinor solutions in the same algebraic form as the usual free-spinor solutions:
\begin{align}
u_L^s &= \sqrt{P_+ \cdot \sigma}\ \xi_s,
&
u_R^s &= \sqrt{P_+ \cdot \bar\sigma}\ \xi_s,
\label{eq:u}
\\
v_L^s &= \sqrt{P_- \cdot \sigma}\ \xi_{-s},
&
v_R^s &= \sqrt{P_- \cdot \bar\sigma}\ \xi_{-s}.
\label{eq:v}
\end{align}
We find that $\dot\theta$ can modify the dispersion relation (between $E_\pm$ and $\vec p$), but not the algebraic form of spinor solutions since $P_\pm^0$ are equal to the original energy $E^{(0)}=\sqrt{|\vec p\,|^{\,2}+m^2}$ when $\thdot=0$ for a given $\vec p$.
Denoting $p^{(0)} = (E^{(0)}, \vec p)$, one can take $P_+ = P_- = p^{(0)}$ in Eqs.\,\eqref{eq:u} and \eqref{eq:v}, and treat $u_{L,R}$ and $v_{L,R}$ as if $\thdot=0$.

\subsubsection{Quantization in the $\thdot$ background}
With the modified dispersion relations arising in the rolling phase field background, it is necessary to examine the quantization rules and the mode expansion of quantum fields.  For a complex scalar $\Phi$ 
or a Dirac fermion $\Psi$ coupling to a rolling phase field $\thdot$,  the mode expansions of their quantum fields are given by 
\begin{align}
\Phi(x) & = \!\! \int \!\! {d^3 \vec p \over  (2\pi)^3 \sqrt{2E^{(0)}}} \left[ b_{\Phi}(\vec p) \, e^{- i p_+ \cdot x} + d_\Phi^\dagger({\vec p}) \, e^{+ i p_- \cdot x} \right], \\
\Psi(x) & =  \!\! \int \!\! {d^3 \vec p \over  (2\pi)^3 \sqrt{2E^{(0)}}} \left[ b_\Psi({\vec p})\, u_{\vec p} \, e^{- i p_+ \cdot x} + d_\Psi^\dagger({\vec p}) \,v_{\vec p}\, e^{+ i p_- \cdot x} \right],
\end{align}
where $p_\pm=(E_\pm,\vec p)$ with $E_\pm=E^{(0)}\mp X\thdot$ for a particle/anti-particle mode with $X=-(B-L)/2$. 
Here, $E^{(0)} \equiv \sqrt{ \vec p^2 +m^2}$ and $X$ ($m$) is the charge (mass) of $\Phi$ or $\Psi$. It should be noted that $E^{(0)}$
 enters the normalization of both particle and antiparticle modes, and this prescription ensures the consistency of the quantization rules:
\begin{align*}
   & [b_\Phi({\vec p}), b^\dagger_\Phi({\vec q})]=[d_\Phi({\vec p}), d^\dagger_\Phi({\vec q})]=(2\pi)^3 \delta^3(\vec p -\vec q),
   \\
  & 
  [\Phi(t,\vec x),\Pi(t,\vec y)]=i\delta^3(\vec x -\vec y), 
\end{align*}
and $[b(\vec p), b(\vec q)]=[d(\vec p),d(\vec q)]=0$ and $[\Phi(t,\vec x), \Phi(t,\vec y)]=[\Phi(t,\vec x), \Phi^\dagger(t,\vec y)]$$=[\Pi(t,\vec x), \Pi(t,\vec y)]=[\Pi(t,\vec x), \Pi^\dagger(t,\vec y)]=0$,
where $\Pi = \dot \Phi^\dagger + i X \thdot \Phi^\dagger$ is the conjugate momenta of $\Phi$. 
In the case of the fermion field $\Psi$, the corresponding property is realized through anticommutator relations; the conjugate momentum is unmodified since $\partial_\mu \theta$ couples without time derivatives, and the normalization follows directly.

\subsection{Decay rates of $\Delta$  }
\label{sec:decay}

\subsubsection{$\Delta \to \ell \ell$}

Let us decompose the interaction between $\Delta$ and leptons
with field expansions 
\begin{align*}
\Delta &\sim b_\Delta  e^{- i p_{\Delta+} \cdot x} + d_\Delta^\dagger e^{+ i p_{\Delta-} \cdot x}, \\
\ell &\sim b_\ell \, u_\ell\, e^{- i p_{\ell+} \cdot x} + d_\ell^\dagger \, v_\ell\, e^{+ i p_{\ell-} \cdot x}, \\
\bar \ell^c &\sim b_\ell \bar v_\ell\, e^{- i p_{\ell+} \cdot x} + d_\ell^\dagger \, \bar u_\ell\, e^{+ i p_{\ell-} \cdot x}.
\end{align*}
We can obtain the squared matrix element of $\Delta \to \ell \ell$ as
\begin{align}
|\mathcal{M}|^2
&=
y^2 |\bar u_{\ell} P_L v_{\ell'}|^2
2 y^2 \left(E^{(0)}_{\ell} E^{(0)}_{\ell'} - \vec{p}_\ell \cdot \vec{p}_{\ell'}\right).
\label{eq:amp_squared_ll}
\end{align}

The energy-momentum conservation follows from the multiplication of $e^{i p\cdot x}$ for each particle and the integration over spacetime, which generates the delta function $\delta^{(4)}(p_\Delta - p_\ell - p_{\ell'})$.
Each $p_i$ follows the dispersion relations \eqref{eq:dispersion_Delta} and \eqref{eq:dispersion_lepton}.
Choosing the rest frame of $\Delta$ where $\vec{p}_\Delta= \vec{p}_\ell+ \vec{p}_{\ell'}=0$, we find $E_{\ell+} = E_{\ell'+} = E_{\Delta+}/2 = (m_\Delta - \thdot)/2$.
Since $E_{\ell +} = E_\ell^{(0)}-\thdot/2$, we obtain $E_{\ell}^{(0)}=E_{\ell'}^{(0)}=m_\Delta/2$, and
\begin{equation}
|\mathcal{M}|^2 =
2y^2\big(2(E_\ell^{(0)})^2 - m^2\big)\simeq 
y^2 m_\Delta^2 ,
\label{eq:amp_squared_ll_2}
\end{equation}
neglecting the lepton mass $m$ in the last step.
Note that Eq.\,\eqref{eq:amp_squared_ll} is essentially the Lorentz invariant product $p^{(0)}_{\ell} \cdot p^{(0)}_{\ell'}$, and thus Eq.\,\eqref{eq:amp_squared_ll_2} is Lorentz invariant.

The partial decay width can be calculated as 
\begin{align}
\Gamma_{\Delta \to \ell\ell}
& =\! \frac  1 2 \frac{1}{2E_\Delta^{(0)}}
\!\! \int \!\! \frac{1}{\!(2\pi)^6\!}\frac{d^3 \vec p_\ell}{2 E_\ell^{(0)}} \! \frac{d^3 \vec p_{\ell'}}{2 E_{\ell'}^{(0)}} \! (2\pi)^{\!4} \! 
\delta^4( {\textstyle \sum}p_{i+}
)|\mathcal{M}|^2
\nn \\
&
=\frac{y^2 m_\Delta }{ 32\pi},
\end{align}
with denoting $\delta^4( {\textstyle \sum}p_{i+})=\delta^4 (p_{\Delta+}-p_{\ell +}-p_{\ell'+})$.
The factor $1/2$ comes from the identical particles in the final states.
Note that the prefactor in the phase space integration takes the form of $1/(2E_i^{(0)})$, not the shifted one, $1/(2E_i)$, following the canonical quantization of the fields.
Inside the delta function, the shifts due to nonzero $\thdot$ compensate (which is a consequence of the fact that $\bar{\ell}^c \ell \Delta^\dagger$ is $(B-L)$-number conserving), and thus the final expression does not include $\thdot$ dependence.
It remains independent of $\thdot$, even when the lepton mass $m$ is retained.

For the decay rate of $\bar \Delta \to \bar \ell \bar \ell$, we can simply take $\thdot \to - \thdot$, and thus $\Gamma_{\bar \Delta \to \bar \ell \bar \ell} = \Gamma_{\Delta \to \ell\ell}$.

\subsubsection{$\Delta \to H H$}

The interaction between $\Delta$ and $H$ is given by $\mu_H \Delta H^\dagger  H^\dagger$.
As the amplitude of $\Delta \to HH$ is simply ${\cal M} = \mu_H$, the partial decay width can be obtained as 
\begin{align}
\hspace{-0.1cm}
\Gamma_{\Delta \to H H}
&= \frac{1}{2} \!\frac{1}{2E_\Delta^{(0)}} \!\!\!
\int \!\!\! \frac{1}{\!(2\pi)^6\!} \frac{d^3 \vec p_H}{2 E_H} \!\frac{d^3 \vec p_{H'}}{2 E_{H'}} \!
(2\pi)^4 
\delta^4({\textstyle \sum} p_{i+} 
)|\mathcal{M}|^2
\nn \\
&
=
\frac{\mu_H^2}{32\pi m_\Delta}.
\end{align}
Note that the dispersion relation of $H$ is not shifted by $\thdot$; $E_H^2 = |\vec p_H|^2+m_H^2$.
As before, the factor $1/2$ is the symmetry factor, and $\delta^4( {\textstyle \sum}p_{i+})=\delta^4 (p_{\Delta+}-p_{H}-p_{H'})$.
Unlike the case of $\Delta \to \ell \ell$, $E_H=E_{H'} = E_\Delta/2 = (m_\Delta -\thdot)/2$ is shifted by $\thdot$.
However, its contribution cancels between the numerator ($p_H^2$) and the denominator ($E_H^2$) in the massless limit, and in the final expression, $\Gamma_{\Delta \to HH}$ becomes independent of $\thdot$.
The replacement $\thdot \to -\thdot$ can be taken to obtain $\bar \Delta \to \bar H \bar H$,
which results in $\Gamma_{\bar \Delta \to \bar H \bar H}=\Gamma_{\Delta \to H H}$ since $\Gamma_{\Delta \to H H}$ is $\thdot$ independent.
Therefore, we use the total width given by
\begin{equation}
\Gamma_\Delta \equiv \Gamma_{\Delta \to \ell\ell} + \Gamma_{\Delta \to H H}
=
\frac{y^2+\mu_H^2/m_\Delta^2}{32\pi}\,m_\Delta.
\end{equation}

There appears no CP asymmetry in both decay modes, $\Delta \to ll$ and $\Delta \to H H$, but the reasons are different. The absence of asymmetry in the first process is attributed to lepton number conservation, whereas in the second process it is a consequence of the phase space approaching the massless limit.  

For clarity, we examine a mother particle characterized by mass $m_A$ and decaying into two daughter particles with respective masses $m_1$ and $m_2$.
Let us assume that the shifted dispersion relations of them are given by $E_i = E^{(0)}_i - \thdot X_i$ for $i=1,2$ and $A$. 
Considering energy conservation, we have
\begin{equation*}
 E_A^{(0)} - \Delta X \dot\theta = E_1^{(0)}+E_2^{(0)},
\end{equation*}
where $\Delta X\equiv X_A-X_1-X_2$. 
One finds the decay rate of the mother particle at rest:
\begin{align}
 \Gamma_A &= {1\over 2 m_A} {|{\vec p}_1| \over \tilde m_A} \int {d\Omega \over 16\pi} |{\cal M}|^2, \\
 {|{\vec p}_1| \over \tilde m_A} &\equiv {1\over2}\sqrt{ \left[ 1- {(m_1+m_2)^2 \over \tilde m_A^2}\right] \left[ 1- {(m_1-m_2)^2 \over \tilde m_A^2}\right]}
\end{align}
where $\tilde m_A \equiv m_A -\Delta X \dot\theta$. In the limit of $m_1^2, m_2^2 \ll \tilde m_A^2$ and $\dot \theta \ll m_A$,  the phase space factor $|{\vec p}_1|/\tilde m_A$ shows asymmetry between the particle and anti-particle decays $A\to 1+2$ and $\bar A \to \bar 1+\bar 2$
\begin{align}
    {|{\vec p}_1| \over \tilde m_A} -  {|{\vec p}_1| \over \tilde m_{\bar A}}
    \approx -2 \Delta X {\dot\theta \over m_A}{m_1^2+m_2^2 \over m_A^2} \, .
\end{align}
Note that CP asymmetry in the Higgs triplet decays appears only in the above phase space as the decay amplitudes of the lepton and Higgs modes are the same for the particles and anti-particles, that is,  $|{\cal M}| = |{\cal \bar M}|$. It is instructive to contrast this with type-I seesaw, where decay amplitudes of a right-handed neutrino depend on $\dot\theta$ \cite{Chun:2025abp}. 

In the early universe under consideration, CP-asymmetric decay occurs due to the thermal mass of the Higgs doublet $m^2_H(T) \approx y_t^2 T^2/4$  as
\begin{align}
    {\Gamma_{\Delta \to H H} - \Gamma_{\bar\Delta \to \bar H \bar H} \over 
    \Gamma_{\Delta \to H H} + \Gamma_{\bar\Delta \to \bar H \bar H}} \approx - y_t^2 {T^2 \over m_\Delta^2} {\dot\theta \over m_\Delta}
\end{align}
for $T \lesssim m_\Delta$, where $\Delta X=1$ for this process. 
We remark that this $CP$-asymmetric decay plays a negligible role, as the Higgs triplet closely follows the equilibrium density till $T\ll m_\Delta$ due to efficient processes of gauge annihilation and inverse decay, which decouple late to determine the final lepton asymmetry as in the strong washout regime of type I seesaw \cite{Chun:2023eqc,Chun:2025abp}.

\section{Boltzmann equations}
\label{sec:boltzmann}

\subsection{Thermal equilibrium shifted by $\thdot$}

In the background of nonzero $\thdot$, the Hamiltonian is shifted as $H = H_0 - \thdot X$, and the density matrix in the grand canonical ensemble becomes 
\bal
\rho = Z^{-1}\exp[-(H_0 - \thdot X - \sum_\alpha \mu_\alpha Q_\alpha)/T],
\label{eq:density_mat}
\eal
where $H_0$ is the Hamiltonian with $\thdot=0$ and $X=-(B-L)/2$ is the number operator coupled to $\thdot$, $Z$ is the partition function, and $\mu_\alpha$ is a chemical potential of conserved quantum number $Q_\alpha$, e.g. hypercharge.
In general, $\mu_\alpha$ are Lagrange multipliers, and their values should be determined by constraint equations.
We can decompose $Q_\alpha$ into the particle number operators, $Q_\alpha = \sum_i (Q_\alpha)_i$ by denoting $(Q_\alpha)_i$ as the $i$ component in the $Q_\alpha$ operator for a particle species $i$.

For simplicity, we assume Maxwell-Boltzmann statistics from this point on, which is a good approximation for our scope.
Defining $\mu_i$ as $\mu_i = \sum_\alpha \mu_\alpha (Q_\alpha)_i$, equilibrium distributions are given by
\bal
f_i^{\rm eq} = \exp[-(E_i^{(0)} - X_i \thdot - \mu_i)/T],
\label{eq:f_i^eq}
\eal
where $X_i$ is the $X$ value of the particle $i$ and is consistent with our dispersion relations.
The equilibrium number density is given by
\bal
n_i^{\rm eq} = g_i \int \frac{d^3 p}{(2\pi)^3} f_i^{\rm eq}
\simeq \bigg(1 + \frac{X_i \thdot + \mu_i}{T} \bigg) n_i^{\rm eq (0)},
\eal
where $n_i^{\rm eq (0)}$ is the number density for $\thdot=0$, and $g_i$ is the degrees of freedom for $i$.
Here, we emphasize that $n_i^{\rm eq}$ is the number density of particles only, and $g_i$ is thus the internal degrees of freedom \emph{without} counting antiparticles. 
Therefore, $g_\Delta = 3$ and $g_j =  2$ for $j=\ell$ and $H$ (in the following discussion, the index $j$ always runs for $\ell$ and $H$).
For the antiparticle $\bar i$, we flip the sign of $\thdot$ and $\mu_i$, i.e. $X_{\bar i} = -X_i$ and $\mu_{\bar i} = -\mu_i$.

When $\thdot = 0$, the constraint equations $\langle Q_\alpha \rangle =0$ together with chemical equilibrium lead to $\mu_i = 0$ for every species, so no asymmetry is generated.
For $\thdot \neq 0$, $\mu_i$ become nonzero and proportional to $\thdot$, but the proportionality coefficients are not simply $X_i$. 
Instead, they are fixed by solving the full system of linear equations enforcing chemical equilibrium. 
This is precisely how spontaneous baryogenesis/leptogenesis works: a rolling background ($\thdot\neq0$) sources chemical potentials for the relevant charges, and chemical equilibrium then converts these into a net baryon/lepton asymmetry.

Let us first obtain $\mu_i$ in Eq.\,\eqref{eq:f_i^eq} in the system of $\Delta, H$ and $\ell$ with only one flavor assuming that both processes $\Delta \leftrightarrow \ell \ell$ and $\Delta \leftrightarrow HH$ are efficient. The only conserved quantity in this system is the hypercharge, whose density is given by
\bal
n_Y = -\Big(n_{\Delta -} + \frac{1}{2} n_{\ell -} + \frac{1}{2} n_{H -}\Big),
\eal
where we define $n_{i -} \equiv n_i - n_{\bar i}$.
While we assume $\ell$ and $H$ are massless, $\Delta$ could be massive, so $n_\Delta^{\rm eq (0)} = g_\Delta \frac{m_\Delta^2 T}{2\pi^2} K_2(m_\Delta/T)$.
The condition $n_Y=0$ then leads to
\bal
&(\thdot + \mu_\Delta)
\frac{g_\Delta}{g_j}\frac{m_\Delta^2}{2T^2}K_2\Big(\frac{m_\Delta}{T}\Big) 
+ \frac{1}{2}\Big(\frac{1}{2}\thdot + \mu_\ell \Big)
+\frac{1}{2}\mu_H =0,
\eal
taking $X_\Delta = 1$, $X_\ell = 1/2$, and $X_H = 0$.
The equilibration conditions of $\Delta \leftrightarrow \ell \ell$ and $HH$, given by $\mu_\Delta = 2\mu_\ell = 2\mu_H$, do \emph{not}
involve $\thdot$.
This is because (i) energy conservation holds in terms of $E_i=E_i^{(0)}- X_i \thdot$ (not $E_i^{(0)}$); $E_\Delta = E_{j1} + E_{j2}$ for $j=\ell$ and $H$, and (ii) the detailed balance from the distribution \eqref{eq:f_i^eq} implies $0=(E_\Delta^{(0)} - X_\Delta \thdot - \mu_\Delta)-(E_{j1}^{(0)}+E_{j2}^{(0)}-2X_j\thdot -2\mu_j)$, where the $E_i^{(0)}$ and $\thdot$ terms vanish.
Therefore, we obtain
\bal
\mu_\Delta = -\frac{1+4r}{2+4r}\thdot,
\quad
\mu_\ell = -\frac{1+4r}{4+8r} \thdot,
\quad
\mu_H = -\frac{1+4r}{4+8r}\thdot,
\eal
where we shorten the equations by defining $$r \equiv {n_\Delta^{\rm eq(0)} \over n_j^{\rm eq(0)}}={1\over2} \frac{g_\Delta}{g_j}\frac{m_\Delta^2}{T^2} K_2({m_\Delta\over T})\, .$$
Note that $r$ goes to $g_\Delta/g_j$
in the massless $\Delta$ limit (i.e. high temperature limit), and 0 for the $T/m_\Delta \to 0$ limit.
Inserting them into Eq.\,\eqref{eq:f_i^eq}, we obtain
\bal
&
f_\Delta^{\rm eq} = \exp[-(E_\Delta^{(0)} - \frac{1}{2+4r}\thdot)/T],
\label{eq:f_Delta^eq}
\\
&
f_\ell^{\rm eq} = \exp[-(E_\ell^{(0)} - \frac{1}{4+8r}\thdot)/T],
\label{eq:f_ell^eq}
\\
&
f_H^{\rm eq} = \exp[-(E_H^{(0)} + \frac{1+4r}{4+8r}\thdot)/T],
\label{eq:f_H^eq}
\eal
and for anti-particles, we flip the sign of $\thdot$.
To simplify the discussion below, we denote $f_i^{\rm eq} = \exp[-(E_i^{(0)} - c_i \thdot)/T]$, i.e. $c_i \equiv X_i + \mu_i/\thdot$.

\subsection{Constructing Boltzmann equations}

We simplify our setup by assuming kinetic equilibrium
\bal
f_i = \bigg( \frac{n_i}{\nEq{i}} \bigg) f^{\rm eq}_i,
\eal
with $f^{\rm eq}_i$ taken from the previous subsection.
Then, the Boltzmann equations become
\bal
&
\dot n_\Delta + 3H n_\Delta
= -\gamma_{\Delta \bar \Delta}
\big(n_\Delta n_{\bar \Delta} - n_\Delta^{\rm eq} n_{\bar \Delta}^{\rm eq} \big)
\nn\\
& \qquad \qquad \qquad
-\gamma_{\Delta \to \ell \ell} \Big(n_\Delta - \frac{n_l^2}{(n_l^{\rm eq})^2} n_\Delta^{\rm eq} \Big)
\nn\\
& \qquad \qquad \qquad
-\gamma_{\Delta \to HH} \Big(n_\Delta - \frac{n_H^2}{(n_H^{\rm eq})^2} n_\Delta^{\rm eq} \Big),
\\
&
\dot n_\ell + 3H n_\ell
=-\gamma_{\ell \bar \ell}
\big(n_\ell n_{\bar \ell} - n_\ell^{\rm eq} n_{\bar \ell}^{\rm eq} \big)
\nn\\
& \qquad \qquad \qquad
+2\gamma_{\Delta \to \ell \ell} \Big(n_\Delta - \frac{n_l^2}{(n_l^{\rm eq})^2} n_\Delta^{\rm eq} \Big),
\\
&
\dot n_H + 3H n_H
=-\gamma_{H\bar H}
\big(n_H n_{\bar H} - n_H^{\rm eq} n_{\bar H}^{\rm eq} \big)
\nn\\
& \qquad \qquad \qquad
+2\gamma_{\Delta \to HH} \Big(n_\Delta - \frac{n_H^2}{(n_H^{\rm eq})^2} n_\Delta^{\rm eq} \Big),
\eal
where $\gamma_{i \bar i}$ are annihilation rates, and $\gamma_{\Delta \to jj} = \frac{K_1(m_\Delta/T)}{K_2(m_\Delta/T)} \Gamma_{\Delta \to jj}$ is the thermal averaged decay rates for $j=\ell$ and $H$.
We omit equations for antiparticles since the structure remains exactly the same as $\gamma_{\Delta \to jj} = \gamma_{\bar \Delta \to \bar j \bar j}$ for $j=\ell$ and $H$.

Let us track the symmetric and antisymmetric number densities; $n_{i\pm} = n_i \pm n_{\bar i}$.
The annihilation rates given by gauge interactions are so large that it is a good approximation to take the symmetric component in equilibrium; $n_{i+} \simeq 2n_i^{\rm eq (0)}$.

For the asymmetric components, it is useful to take
\bal
\frac{n_j^2}{(n_j^{\rm eq})^2}n_\Delta^{\rm eq}
-
\frac{n_{\bar j}^2}{(n_{\bar j}^{\rm eq})^2}n_{\bar \Delta}^{\rm eq}
\simeq
2n_{\Delta}^{\rm eq(0)} 
\Big( \frac{n_{j-}}{n_j^{\rm eq(0)}}
+
(c_\Delta -2c_j) \frac{\thdot}{T}
\Big),
\eal
for $j=\ell$ and $H$, ignoring $\thdot^2$ and higher order terms.
Then, Boltzmann equations for the asymmetric components are given by
\bal
&
\dot n_{\Delta-} \!\! + 3H n_{\Delta-} \!\! 
= 
-\gamma_{\Delta \to \ell \ell} \bigg(n_{\Delta-} \!\! - 2n_{\Delta}^{\rm eq(0)} \frac{n_{\ell -}}{n_\ell^{\rm eq(0)}} \bigg)
\\
& \qquad \qquad \qquad
-\gamma_{\Delta \to HH} \bigg(n_{\Delta-} \!\! - 2n_\Delta^{\rm eq(0)} \Big(\frac{n_{H-}}{n_H^{\rm eq(0)}} + \frac{\thdot}{T} \Big) \bigg),
\nn\\
&
\dot n_{\ell-} \!\! + 3H n_{\ell-} \!\! 
= 2\gamma_{\Delta \to \ell \ell} \bigg(n_{\Delta-} \!\! - 2n_{\Delta}^{\rm eq(0)} \frac{n_{\ell -}}{n_\ell^{\rm eq(0)}} \bigg),
\\
&
\dot n_{H-} \!\! + 3H n_{H-} \!\! 
= 
2\gamma_{\Delta \to HH} \bigg(n_{\Delta-} \!\! - 2n_\Delta^{\rm eq(0)} \Big(\frac{n_{H-}}{n_H^{\rm eq(0)}} + \frac{\thdot}{T} \Big)
\bigg).
\eal
Note that the source terms all vanish when $m_\Delta/T \gg 1$ since $n_\Delta^{\rm eq(0)} \propto e^{-m_\Delta/T}$.
This is because we only consider the source term by the decay of $\Delta$.
The next leading-order source term would be an off-shell $\Delta$-mediated scattering process, which is equivalent to considering the Weinberg operator\,\cite{Weinberg:1979sa}.
It is inefficient cosmologically for $T\lesssim 10^{13}\,\GeV$, and thus we do not include it in this work.
The hypercharge conservation can be checked explicitly from the above equations; the Boltzmann equation for $n_Y = -n_{\Delta -} - \frac{1}{2}(n_{\ell -} + n_{H-})$ results in a vanishing collision term.

We define the yield of each component by $Y = n/s$ with the entropy density $s$ and take $z\equiv m_\Delta/T$ as the variable for the time evolution.
With approximating $dT/dt \simeq -HT$, i.e. $dz/dt \simeq z H$ and assuming radiation domination, we obtain
\bal
& \frac{dY_{\Delta -}}{d\ln z} 
 = 
- {\rm Br}_{\ell}  \, \gamma_{\rm D}
\Big(Y_{\Delta-}  -  2r Y_{\ell -}  \Big)
\label{eq:Y_Delta-}
\\
& \qquad 
- {\rm Br}_{H}  \, \gamma_{\rm D}
\Big(Y_{\Delta-} -  2r Y_{H-}  - 2 \frac{\epsilon_1}{z^2} Y_\Delta^{\rm eq (0)}\Big),
\nn \\
&
\frac{dY_{\ell -}}{d\ln z}
= 
2 {\rm Br}_{\ell}  \, \gamma_{\rm D} 
\Big(Y_{\Delta-} - 2r Y_{\ell -} \Big),
\label{eq:Y_ell-}
\\
&\frac{dY_{H -}}{d\ln z}
 = 
2 {\rm Br}_{H}  \, \gamma_{\rm D} \Big(Y_{\Delta-} -  2r Y_{H-} - 2 \frac{\epsilon_1}{z^2} Y_\Delta^{\rm eq (0)} \Big),
\label{eq:Y_H-}
\eal
where ${\rm Br}_j = \Gamma_{\Delta \to jj}/\Gamma_\Delta$,   
\bal
&\gamma_{\rm D} \equiv K \,  z^2 \frac{K_1(z)}{K_2(z)} ,
\eal
and
\bal
K=\Big(\frac{\Gamma_\Delta}{H}\Big)_{T=m_\Delta},
\qquad
\epsilon_1=\Big(\frac{\thdot}{T}\Big)_{T=m_\Delta}.
\eal
Remind that $r\equiv \frac{1}{2}z^2K_2(z)$ is a function of $z$.
Here, we take $\thdot \propto T^3$ assuming the radial mode of $\Phi$ is unchanged, and replace $\thdot/T = \epsilon_1/z^2$.
This relation can change if the radial mode evolves with temperature change (see, e.g., Ref.\,\cite{Chun:2024gvp}, where $\thdot/T \sim {\rm const}$ can be realized in a certain range).

The complete Boltzmann equations \eqref{eq:Y_Delta-}--\eqref{eq:Y_H-} contain the source term with the Boltzmann suppression factor coming from $Y_\Delta^{\rm eq(0)}$, while the equilibration process is controlled by the $r$ term. In the appropriate limits, these equations reproduce the results of Refs.\,\cite{Chun:2023eqc, Berbig:2025hlc}. By solving the full set of Boltzmann equations, we consistently track the generation and evolution of all asymmetries, as discussed in the following section.
The final value of $Y_{\ell -}$ is not expected to deviate significantly from the qualitative estimates, since the freeze-out is primarily determined by the late decoupling of the inverse decay processes.
However, this correction is crucial in the evolution of $Y_{\Delta-}$.

\begin{figure}[t]
\includegraphics[width=1\linewidth]{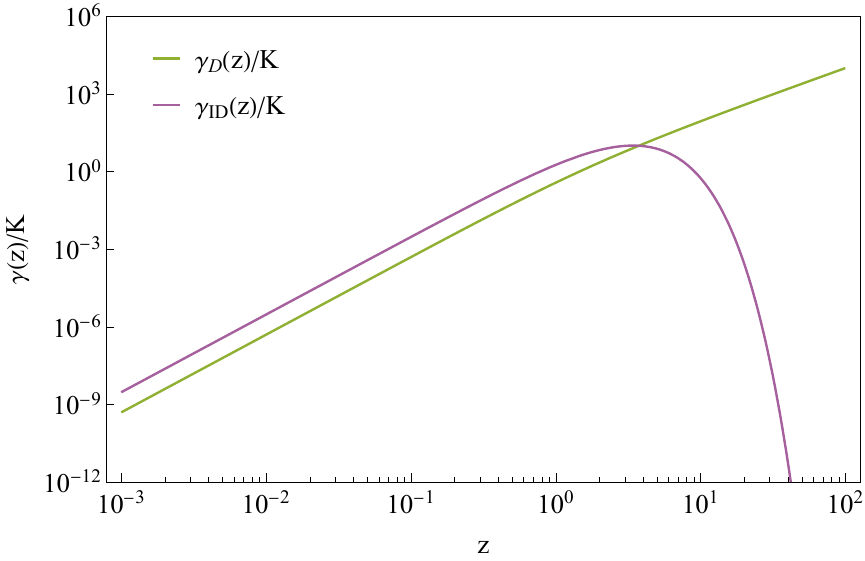}
    \caption{Behavior of the functions $\gamma_{\rm D}(z)/K$ (green) and $\gamma_{\rm ID}(z)/K$ (purple).}
    \label{fig:gammaD}
\end{figure}

Let us now find the relaxation points indicated by each term.
The relaxation rates for terms in $Y_{\Delta -}$ are simply given by ${\rm Br}_j \gamma_D$, and those for $Y_{j-}$ can be identified by ${\rm Br}_j \gamma_{ID}$ 
with the definition,
\bal
\gamma_{\rm ID} \equiv 
4r \, \gamma_D
= 2 \frac{g_\Delta}{g_j} K \,  z^4 K_1(z).
\eal
In Fig.\,\ref{fig:gammaD}, we depict $\gamma_{\rm D}/K$ and $\gamma_{\rm ID}/K$.
Note that $\gamma_{\rm ID}(z)$ gets Boltzmann suppressed at $z \gtrsim 1$ while $\gamma_{\rm D}(z)$ increases even after $z\gtrsim 1$.

Then, the relaxation point can be summarized as
\bal
& {\rm Br}_\ell \,\gamma_{\rm D}>1: & Y_{\Delta -} & \! \to 2r Y_{\ell -} ,
\label{eq:relax_Delta_l}
\\
& {\rm Br}_H \,\gamma_{\rm D}>1: & Y_{\Delta -} & \! \to 2r Y_{H -} + 2 \frac{\epsilon_1}{z^2}Y_\Delta^{\rm eq(0)} ,
\label{eq:relax_Delta_H}
\\
& {\rm Br}_\ell \,\gamma_{\rm ID}>1: & Y_{\ell -} &\! \to \frac{Y_{\Delta -}}{2r} ,
\label{eq:relax_l}
\\
& {\rm Br}_H \,\gamma_{\rm ID}>1: & Y_{H -} & \! \to \! \frac{Y_{\Delta-}}{2r} \! - \! \frac{\epsilon_1}{r\, z^2}Y_\Delta^{\rm eq(0)} .
\label{eq:relax_H}
\eal
Assuming all the relaxation rates are greater than one, the relaxation becomes 
\bal
Y_{\Delta-} &\to Y_{\Delta-}^{\rm eq}=\frac{1}{1+2r} \frac{\epsilon_1}{z^2} Y_{\Delta}^{\rm eq(0)},
\label{eq:Y_Delta-^eq}
\\
Y_{\ell-} &\to Y_{\ell-}^{\rm eq}=\frac{1}{2+4r} \frac{\epsilon_1}{z^2} Y_{\ell}^{\rm eq(0)},\label{eq:Y_l-^eq}
\\
Y_{H-} &\to  Y_{H-}^{\rm eq}=-\frac{1+4r}{2+4r} \frac{\epsilon_1}{z^2} Y_{H}^{\rm eq(0)},\label{eq:Y_H-^eq}
\eal
which are nothing but the equilibrium values indicated from Eqs.\,\eqref{eq:f_Delta^eq}--\eqref{eq:f_H^eq} 
while $\thdot/T \equiv \epsilon_1/z^2$.
However, when one of the branching ratios is small, the yields do not reach the equilibrium values while some of the relaxation conditions are satisfied. 
We present a detailed case-by-case study in the next section, using numerical result.

It is noteworthy that $\epsilon_1$ dependence disappears in the Boltzmann equations if we rewrite Eqs.\,\eqref{eq:Y_Delta-}--\eqref{eq:Y_H-} for $\tilde Y_{i-} \equiv Y_{i-}/\epsilon_1$.
We thus solve the Boltzmann equations numerically in terms of $\tilde Y_{i-}$, and one can obtain the required $\epsilon_1$ to explain the observed baryon asymmetry.
Although we have not included the quark sector in our Boltzmann network, the final $Y_{\ell -}$ should be understood as a $B-L$ number generation $Y_{B-L}=c_{B-L} Y_{l-}$ where $c_{B-L}$ is influenced by the flavor effect, which depends on the leptogenesis temperature $T\sim m_\Delta$.
In the flavor-blind regime of $T\lesssim 10^5$ GeV\,\cite{Bodeker:2019ajh}, when the electron Yukawa coupling finally comes into thermal equilibrium, we get $c_{B-L}=-158/11$, considering the generation of the lepton and Higgs asymmetry satisfying $Y_{H-}+Y_{l-}=0$ in the type-II leptogenesis. 
Then, the final baryon number is obtained as
\bal
Y_B^{\rm (obs)} = \frac{28}{79} c_{B-L} \epsilon_1 \tilde Y_{\ell -}^\infty,
\label{eq:YB_obs}
\eal
where $\tilde Y_{\ell -}^\infty$ denotes the final $\tilde Y_{\ell -}$ obtained by solving the Boltzmann equations, and the observed baryon asymmetry is $Y_B^{\rm (obs)} = (8.7 \pm 0.1)\times 10^{-11}$\,\cite{Planck:2018vyg}.

\section{Numerical results}
\label{sec:results}

Our model parameters can be fixed by choosing physical quantities $(m_\nu)_{\alpha \beta}$, ${\rm Br}_\ell$, and $m_\Delta$.
In our study, we make a rough conversion by taking a single-flavored lepton and defining the effective neutrino mass $\tilde m_\nu = y \mu_H v_{\rm EW}^2/m_\Delta^2$.
In this approximation, the $m_\Delta$ dependence also disappears since its effect in Boltzmann equations only appears in the washout factor $K$, which is completely determined when ${\rm Br}_\ell$ and $m_\nu$ are fixed;
\bal
K
&=
\frac{\Gamma_\Delta}{H_1}
=\frac{\frac{y^2+\mu_H^2/m_\Delta^2}{32\pi}m_\Delta}{\sqrt{\frac{\pi^2 g_*}{90} }\frac{m_\Delta^2}{M_{\rm Pl}}}
=
\sqrt{\frac{90}{\pi^2 g_*}}
\frac{ \tilde m_\nu M_{\rm Pl}}{32\pi v_{\rm EW}^2}
\frac{1}{\sqrt{{\rm Br}_\ell {\rm Br}_H}}
\nn \\
&\simeq
\frac{12}{\sqrt{{\rm Br}_\ell {\rm Br}_H}}
\bigg(\frac{\tilde m_\nu}{0.05\ \text{eV}}\bigg)
\bigg(\frac{106.75} {g_*}\bigg)^{\!\!1/2},
\eal
using the relation $y^2+\mu_H^2/m_\Delta^2 = \frac{1}{\sqrt{{\rm Br}_\ell {\rm Br}_H}} \frac{\tilde m_\nu m_\Delta}{v_{\rm EW}^2}$.
Here, $M_{\rm Pl}= 2.44\times 10^{18}\,\GeV$ is the reduced Planck mass,
and we take the effective relativistic degrees of freedom $g_*\simeq 106.75$ as a benchmark value, whose dependence on the final result is mild.
We also fix the effective neutrino mass $\tilde m_\nu = 0.05\,\eV$ to be the scale of atmospheric neutrino mass difference $\Delta m_{\rm atm}^2 \simeq (0.05\,\eV)^2$\,\cite{ParticleDataGroup:2026mpi}.

Note that $m_\Delta$ dependence does not appear in our Boltzmann equations \eqref{eq:Y_Delta-}--\eqref{eq:Y_H-}, and thus our final result is independent of $m_\Delta$ unless it is lighter than a few times the electroweak sphaleron decoupling temperature.
However, it can actually change the efficiency of the annihilation rate $\gamma_{\Delta \bar \Delta}/H$, and the symmetric yield may get some departure from equilibrium. 
We numerically check that this effect is negligible, and thus we simply neglect its effect in the following.
With this setup, the remaining part is the investigation along ${\rm Br}_\ell$.

\begin{figure*}[t]
\includegraphics[width=0.3\textwidth]{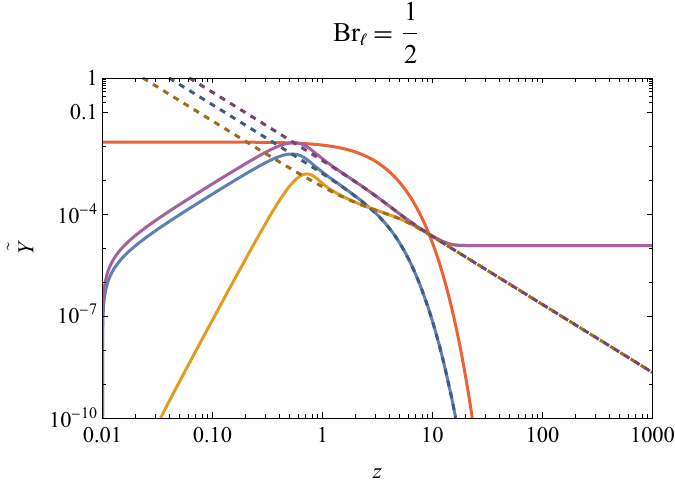}
\includegraphics[width=0.3\textwidth]{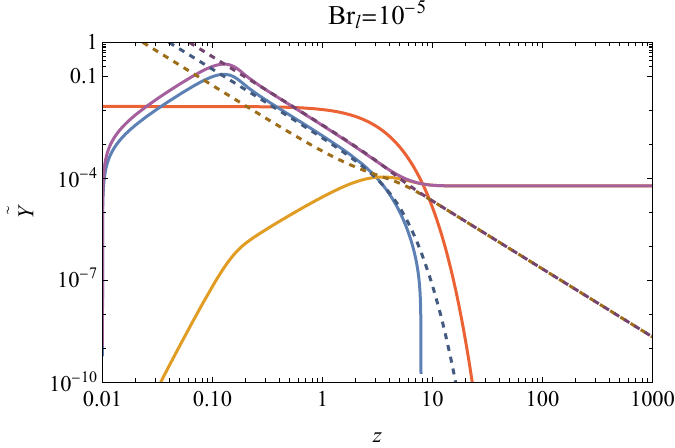}
\includegraphics[width=0.38\textwidth]{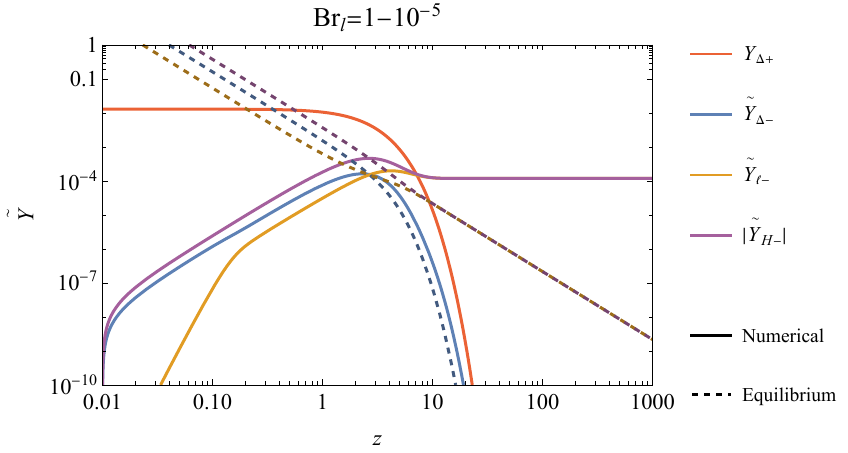}
\caption{
Evolution of $Y_{\Delta +}$ (red), $\tilde Y_{\Delta -}$ (blue), $\tilde Y_{\ell -}$ (orange), and $\tilde Y_{H -}$ (purple) as functions of $z$. 
The solid lines correspond to the numerical solutions of Boltzmann equations~\eqref{eq:Y_Delta-}--\eqref{eq:Y_H-}, while dashed lines depict equilibrium values given in Eqs.\,\eqref{eq:Y_Delta-^eq}--\eqref{eq:Y_H-^eq}.
We take ${\rm Br}_\ell = 0.5$ (left), $10^{-5}$ (middle), and $1-10^{-5}$ (right).
}
\label{fig:evol}
\end{figure*}

Let us first consider the case when ${\rm Br}_\ell $ and ${\rm Br}_H$ are comparable.
In this case, $K\sim 24$, and all three components, $Y_{\Delta -}$, $Y_{\ell -}$ and $Y_{H-}$, reach the thermal equilibrium. 
Later, the inverse decay decouples, due to the exponential suppression, and thus $Y_{\ell-}$ and $Y_{H-}$ are frozen.
The freeze-out value of $Y_{\ell -}$ is determined by the decoupling of species having the smaller branching ratio due to hypercharge conservation; $Y_{\ell -} = - Y_{H-}$ at $z_{\rm fo} \simeq 12 + \frac{1}{2} \log[{\rm Br}_</(1-{\rm Br}_<)]$ with ${\rm Br}_< = \min ({\rm Br}_\ell, \, {\rm Br}_H)$.
Since it is a freeze-out process, the final value of $Y_{\ell-}$ is given by $Y_{\ell-}^{\rm eq}$ at $z_{\rm fo}$, and it is roughly $O(10^{-5})  (10/z_{\rm fo})^2\epsilon_1$
(see also the derivation of $z_{\rm fo}$ presented in Appendix.\,\ref{app:analytic}).
This scenario is numerically shown in the left panel of Fig.\,\ref{fig:evol}, where we depict the evolution of $Y_{\Delta+}$ (red), $Y_{\Delta-}$ (blue), $Y_{\ell-}$ (orange) and $Y_{H-}$ (purple) as functions of $z$ for ${\rm Br}_\ell ={\rm Br}_H=1/2$.

When ${\rm Br}_\ell \ll 1$, the relaxation conditions \eqref{eq:relax_Delta_l} and \eqref{eq:relax_l} are always broken, while the others can still hold for a range of $z$.
Thus, the final $Y_{\ell-}$ value is determined by the freeze-in process;
\bal
Y_{\ell -}^\infty \simeq \int_{z_0}^{z_f} d \ln z \, 2 {\rm Br}_\ell \gamma_D Y_{\Delta -},
\label{eq:freeze_in}
\eal
where $Y_{\Delta -} \simeq \frac{2}{1+4r}\frac{\epsilon_1}{z^2}Y_\Delta^{\rm eq(0)}$ is slightly shifted from Eq.\,\eqref{eq:Y_Delta-^eq} due to the absence of $Y_{\ell-}$ in the thermal equilibrium,
and $z_0$ and $z_f$ are initial and final values of $z$ in the estimation.
This results in $O(10^{-2})\sqrt{{\rm Br}_\ell}$ (see Appendix.\,\ref{app:analytic} for derivation), and can be seen in our numerical result presented in the middle panel of Fig.\,\ref{fig:evol}; $Y_{\ell-}$ freezes in, $Y_{H-}$ freezes out and their frozen values match at large $z$.
On the other hand, $Y_{H-}$ reaches its relaxation point \eqref{eq:relax_H}
and then freezes out at $z_{\rm fo}^H$, which is greater than the freeze-in value of $Y_{\ell-}$.
During this procedure, the relaxation point is now slightly tilted by nonzero $Y_{\ell-}$, to conserve the hypercharge.
Finally, we have $Y_{\ell -} = -Y_{H-}$ at large $z$.

If $1- {\rm Br}_\ell \ll 1$ (i.e., ${\rm Br}_H \ll 1$), the relaxation rates of $\Delta \to HH$ and its inverse process never reach one.
Meanwhile, $Y_{\ell -}$ reaches its relaxation point $\sim Y_{\Delta -}/(z^2 K_2(z))$ indicated in Eq.\,\eqref{eq:relax_l}.
However, the actual source term driven by the nonzero $\thdot$ only appears in the $\Delta \leftrightarrow HH$ mode, and therefore, all three components undergo a freeze-in process (see the right panel of Fig.\,\ref{fig:evol}).
Therefore, the final abundance can be estimated by 
\bal
Y_{\ell -}^\infty \simeq 
-\left. Y_{H -} \right|_{z=z_{\rm fo}}
\simeq
\int_{z_0}^{z_{\rm fo}}
d\ln z \, 4{\rm Br}_H \gamma_D
\frac{\epsilon_1}{z^2} Y_{\Delta}^{\rm eq(0)}.
\eal
Similarly to the previous case, this results in $\tilde Y_{\ell -} \simeq O(10^{-1})\sqrt{ {\rm Br}_H}$
(a derivation is given in Appendix.\,\ref{app:analytic}).

\begin{figure}[t]
\includegraphics[width=0.45\textwidth]{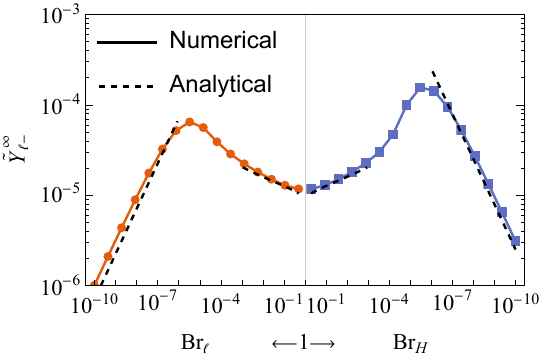}
    
    \caption{The final asymmetry $\tilde Y_{\ell-}^\infty$ with ${\rm Br}_\ell$. 
    }
    \label{fig:Yfinal}
\end{figure}

Altogether, we show our numerical result varying ${\rm Br}_{\ell}$ in Fig.\,\ref{fig:Yfinal}.
As expected, $\tilde Y^\infty_{\ell-} \sim 2\times 10^{-5}$ when ${\rm Br}_\ell$ and ${\rm Br}_H$ are comparable, while $Y^\infty_{\ell -} \propto \sqrt{{\rm Br}_\ell {\rm Br}_H}$ if either of ${\rm Br}_\ell$ or ${\rm Br}_H$ is small.
Our result show that the $\tilde Y_\ell^\infty \lesssim 2\times 10^{-4}$, implying that $|\epsilon_1| \gtrsim 9\times 10^{-8}$ to explain the observed baryon asymmetry as indicated in Eq.\,\eqref{eq:YB_obs}.

\section{Relic majoron abundance}
\label{sec:cogenesis}

After leptogenesis, $\thdot$ keeps being redshifted, gets eventually trapped by the majoron potential, and it oscillates.
The majoron oscillation behaves like matter, so let us estimate its abundance.

Given $\epsilon_1$ for successful leptogenesis, one obtains the majoron number density $n_\theta= \thdot  v_\sigma^2=\epsilon_1 m_\Delta v_\sigma^2 $ at $T=m_\Delta$, or equivalently, $Y_\theta\equiv n_\theta/s \approx 0.02 \epsilon_1 v^2_\sigma/m^2_\Delta $. 
Assuming a small constant majoron mass $m_J$ driven by a soft-breaking operator, 
the majoron relic can be estimated as $\frac{\rho_J}{s} \simeq 2 m_J Y_\theta$ from the kinetic misalignment\,\cite{Co:2019jts}. 
Since it has to be less than the observed dark matter relic $\frac{\rho_{\rm DM}^{\rm obs}}{s} \simeq 0.44\,\eV$, we have
\begin{align}
   v_\sigma &< \frac{3.2 \,\TeV}{\sqrt{|\epsilon_1|}}  \bigg( {m_\Delta \over \TeV} \bigg) \sqrt{\eV \over m_J} 
   \\ 
   &\lesssim 1.1\times 10^7 \GeV \bigg( {m_\Delta \over \TeV}\bigg) {\sqrt{\eV \over m_J}} , \nn
\end{align}
where we used $|\epsilon_1| \gtrsim 10^{-7}$. 
This constraint may be avoided if the majoron decays before recombination, depending on its lifetime and abundance\,\cite{Chang:2024mvg}.

If the majoron consists of the whole dark matter, it has a constraint on its lifetime $\tau_J > 250\,{\rm Gyr}$ from the CMB and Baryon Acoustic Oscillations (BAO) analysis\,\cite{Audren:2014bca, Enqvist:2019tsa, Nygaard:2020sow, Alvi:2022aam, Simon:2022ftd}, whose excluded region can be seen in Fig.~3 of Ref.\,\cite{Chun:2023eqc} or Fig.~6 and 7 of Ref.\,\cite{Berbig:2025hlc}.
For the majoron mass above the $\MeV$ scale, strong constraints\,\cite{Akita:2023qiz} come from measurements of neutrino flux\,\cite{Borexino:2019wln, KamLAND:2021gvi, Olivares-DelCampo:2017feq, Palomares-Ruiz:2007egs, Frankiewicz:2016nyr, Super-Kamiokande:2011lwo, Super-Kamiokande:2021jaq, IceCube:2021kuw, IceCube:2023ies, Arguelles:2022nbl, Albert:2016emp}.

\section{Conclusion}
\label{sec:conclusion}

We have analyzed majoron-driven spontaneous leptogenesis in the type-II seesaw model. 
The model is minimal: a single $SU(2)_L$-triplet $\Delta$ and a $U(1)_{B-L}$-charged singlet $\Phi$, whose vev simultaneously generates the trilinear coupling $\mu_H HH\Delta^\dagger$ and provides the majoron as its angular mode. 
Neutrino masses arise as usual from the induced triplet vev, and no second triplet, no additional heavy state, and no CP-violating phase beyond the rolling background itself are required.

The mechanism operates differently from the thermal leptogenesis scenario. 
Working in the basis where the majoron phase is removed from the interactions, we found that the resulting shifts of the dispersion relations do not generate any asymmetry between $\Gamma_{\Delta \to jj}$ and $\Gamma_{\bar \Delta \to \bar j \bar j}$. 
What $\thdot$ does instead is to shift the equilibrium state itself. 
The distribution functions in the rolling background carry chemical potentials proportional to $\thdot$ fixed by hypercharge neutrality together with the chemical-equilibrium conditions of the two decay channels, and the asymmetry is generated as the decays and inverse decays drive the plasma toward this shifted equilibrium. 
Consequently, the obstacles in the thermal leptogenesis case are no longer problematic in our scenario; the lepton asymmetry is not suppressed by the large gauge annihilation rate, and the strong washout factor enforced by the neutrino mass scale is what makes the mechanism more efficient.

Solving the Boltzmann equations for the asymmetric yields, we found that the final asymmetry is essentially independent of the triplet mass once $\epsilon_1$ is fixed, and is controlled almost entirely by the branching ratio ${\rm Br}_\ell$. 
When both channels are comparable, all three asymmetries reach their shifted equilibrium values and freeze out when the inverse decay of the weaker channel decouples. 
When either branching ratio is hierarchically small, the corresponding asymmetries instead freeze in, and the final yield is suppressed as $\sqrt{{\rm Br}_\ell {\rm Br}_H}$. 
This suppression has a simple symmetry origin: if either coupling in Eq.\,\eqref{eq:lagrangian} is switched off, lepton number is restored as an independent conserved charge, the majoron phase becomes removable by a field redefinition, and no asymmetry can be generated however fast the field rolls. 
Both decay channels are therefore indispensable, and our numerical results confirm this expectation across ten orders of magnitude in ${\rm Br}_\ell$ and ${\rm Br}_H$.

Several simplifications deserve comment. 
We have restricted the Boltzmann network to $\Delta$, a single lepton flavor, and $H$, omitting the remaining SM species and the quark sector; including them modifies the chemical-potential relations and hence the final yield by a factor of order unity. 
We have treated the radial mode as frozen at $v_\sigma$, so that $\thdot \propto T^3$; if instead it evolves, the temperature dependence of the source changes and the analysis must be redone with the appropriate scaling. 
We have also neglected the off-shell $\Delta$-mediated scattering that would source the asymmetry at temperatures where $\Delta$ is Boltzmann-suppressed, which is justified for $T \lesssim 10^{13} \GeV$.

Finally, since the asymmetry is essentially independent of $m_\Delta$, it is natural to ask how low the triplet mass may be pushed. 
The relevant requirement is that the source be active before the electroweak phase transition, which leaves room well below the high-scale benchmark adopted here and, in principle, down to the region probed by current collider searches for doubly charged scalars. 
A dedicated study of this regime, together with the majoron phenomenology such as its contribution to $N_{\rm eff}$ would be worthwhile. 
We leave these to future work.

\begin{acknowledgments}
This work of THJ and JS was supported by IBS, under the project code IBS-R018-D1. 
\end{acknowledgments}

\begin{appendix}

\section{Analytic understanding of final asymmetry}
\label{app:analytic}

In this appendix we provide analytic estimates of the final asymmetry in the three regimes identified in Sec.\,\ref{sec:results}, and show that they reproduce our numerical results.

\noindent
{\bf \underline{(Case I)} Mild hierarchy (${\rm Br}_\ell \gtrsim 10^{-2}$ and ${\rm Br}_H \gtrsim 10^{-2}$):}
In this case all three components reach their equilibrium values, so that the final asymmetry is set by freeze-out. 
Since $Y_{\ell-}$ is sourced only indirectly, through $Y_{\Delta-}$, its evolution stops as soon as either inverse process decouples, and the earlier of the two decouplings therefore determines the freeze-out. 
If $HH \to \Delta$ decouples first, no further asymmetry is fed into the system and $Y_{\ell-}$ is frozen; if $\ell \ell \to \Delta$ decouples first, $Y_{\ell-}$ is frozen because it can no longer follow the source, even though the source itself is still active. 
Either way the freeze-out is controlled by the channel with the smaller branching ratio, and the freeze-out time $z_{\rm fo}$ follows from
\bal
1 = \Big({\rm Br}_< \, \gamma_{\rm ID} \Big)_{z=z_{\rm fo}}
=2\frac{g_\Delta}{g_j} {\rm Br}_< K z_{\rm fo}^4 K_1(z_{\rm fo})
\label{eq:freeze_out_condition}
\eal
where we define ${\rm Br}_< = \min ({\rm Br}_\ell, \, {\rm Br}_H)$.

Eq.\,\eqref{eq:freeze_out_condition} does not have a solution when ${\rm Br}_< K \, \gtrsim 1$ (i.e., ${\rm Br}_< \gtrsim 10^{-2}$). 
It has two solutions in the opposite case, and we denote the greater solution as $z_{\rm fo}$, corresponding to the moment when the system exits thermal equilibrium.
Here, since the process undergoes the freeze-out process, the final abundance is determined solely by $z_{\rm fo}$ which can be approximated as
\bal
z_{\rm fo} &\simeq 
\log \bigg[12
\frac{g_\Delta}{g_j} \sqrt{\frac{2\pi {\rm Br}_<}{1-{\rm Br}_<}}  (z_{\rm fo}^{\rm ref})^{7/2} \bigg]
\\
&\simeq
12 + \frac{1}{2} \log \bigg[
\frac{{\rm Br}_<}{1-{\rm Br}_<} \bigg],
\eal
with $z_{\rm fo}^{\rm ref}\simeq 11$.
A more asymmetric pair of branching ratios thus makes the freeze-out occur earlier, and the final asymmetry is correspondingly enhanced.

The final asymmetry can be estimated as
\bal
\tilde Y_{\ell -}^\infty \sim  O(10^{-5}) \Big( \frac{10}{z_{\rm fo}} \Big)^2,
\eal
taking $Y_\ell^{\rm eq(0)}\sim0.005$,
which agrees well with our numerical result if we take the prefactor $1.5\times 10^{-5}$.

{\bf \underline{(Case II)} ${\rm Br}_\ell \ll 10^{-2}$:}
Here the $\ell$ channel is too weak for $Y_{\ell-}$ to reach its relaxation value, and its final abundance is instead determined by the freeze-in integral of Eq.\,\eqref{eq:freeze_in}. 
The triplet asymmetry, on the other hand, is still driven efficiently by the $HH$ channel and follows its relaxation value. Denoting by $z_{\rm in}$ the value of $z$ at which $Y_{\Delta-}$ becomes fully relaxed, we approximate
\bal
Y_{\ell -}^\infty \sim 10^{-3} {\rm Br}_\ell \int_{z_{\rm in}}^{O(1)}
d \ln z \,  \gamma_D  \frac{\epsilon_1}{z^2},
\eal
where we approximate $Y_{\Delta }^{\rm eq(0)}\sim 10^{-3}$ 
for $z \lesssim O(1)$, while, at larger $z$, it gets suppressed by the Boltzmann factor.
Taking $\gamma_D \simeq K z^3/2$ in this range $z \lesssim O(1)$, we obtain
\bal
\tilde Y_{\ell -}^\infty \sim O(10^{-2}) \sqrt{{\rm Br}_\ell}.
\label{eq:Y_ell-_case2}
\eal
It agrees well with our numerical results when the prefactor is taken to be $7\times 10^{-2}$.

Meanwhile, the inverse decay $HH \to \Delta$ remains efficient, so that $Y_{H-}$ undergoes a freeze-out instead, and one might expect its final value to be given by its relaxation value at the freeze-out point $z_{\rm fo}^H$. 
This estimate requires care: once $Y_{\ell-}$ has frozen in, the relaxation value of $Y_{H-}$ is no longer the naive one but is tilted by this nonzero constant. 
Hypercharge conservation makes this immediate, since the relaxation point must satisfy $Y_{H-} \to -Y_{\ell-} - 2Y_{\Delta-}$. 
As the freeze-out of $H$ occurs after the freeze-in of $\ell$, the two are locked together and we recover Eq.\,\eqref{eq:Y_ell-_case2}. 
The freeze-in description and estimate are therefore the correct one.

{\bf \underline{(Case III)} ${\rm Br}_H \ll 10^{-2}$:}
This case is the opposite to the previous one: $Y_{H-}$ and $Y_{\Delta-}$ freeze in, while $Y_{\ell-}$ follows its relaxation value. 
Since hypercharge conservation gives $Y_{\ell-}^\infty = -Y_{H-}^\infty$ at large $z$, it can be estimated as 
\bal
\tilde Y_{\ell -}^\infty &\sim 10^{-3} {\rm Br}_H \int_{z_0}^{O(1)}
d \ln z \,  \gamma_D  \frac{1}{z^2}
\nn \\
&\sim O(10^{-1}) \sqrt{{\rm Br}_H}.
\eal
The numerical results imply the prefactor to be $\sim 0.25$.

Combining the three cases, the final asymmetry scales as $\sqrt{{\rm Br}_\ell {\rm Br}_H}$ whenever either branching ratio is small, as anticipated and confirmed by Fig.\,\ref{fig:Yfinal}.

\end{appendix}

\bibliography{references}

\end{document}